\documentclass[a4paper,aps,pra,reprint,superscriptaddress,floatfix]{revtex4-1}
\usepackage{amsmath,amssymb,amsfonts}
\usepackage{algorithmic}
\usepackage{graphicx}
\usepackage{textcomp}
\usepackage[colorlinks,
	linkcolor=red,
	citecolor=blue,
	urlcolor=red]{hyperref}
\usepackage[all]{hypcap}
\usepackage{cleveref}
\newcommand{\coop}{\mathcal{C}}

\begin{document}

\title{Nonreciprocity in microwave optomechanical circuits\footnote{prepared}}

\author{N.~R.~Bernier} 
\author{L.~D.~T\'{o}th}
\author{A.~K.~Feofanov}
\email{alexey.feofanov@epfl.ch}
\affiliation{Institute of Physics, {\'E}cole Polytechnique F{\'e}d{\'e}rale de Lausanne, 
	Lausanne 1015, Switzerland}
\author{T.~J.~Kippenberg}
\email{tobias.kippenberg@epfl.ch}
\affiliation{Institute of Physics, {\'E}cole Polytechnique F{\'e}d{\'e}rale de Lausanne, 
	Lausanne 1015, Switzerland}


\begin{abstract}
  Nonreciprocal devices such as isolators and circulators
  are necessary to protect sensitive apparatus from unwanted noise.
  Recently, a variety of alternatives were proposed
  to replace ferrite-based commercial technologies,
  with the motivation to be integrated with 
  microwave superconducting quantum circuits.
  Here, we review  isolators 
  realized with microwave optomechanical circuits
  and present a gyrator-based picture to develop an intuition
  on the origin of nonreciprocity in these systems.
  Such nonreciprocal optomechanical schemes show promise
  as they can be extended to circulators and directional amplifiers,
  with perspectives to reach the quantum limit in terms of added noise.
\end{abstract}
\date{\today}

\maketitle

\section{Introduction}
Nonreciprocal devices, such as isolators, circulators, and directional amplifiers, are pivotal components for quantum information processing with superconducting circuits because they protect sensitive quantum states from readout electronics backaction. 
The commercial Faraday-effect devices typically used in these applications
have a large size and weight that require bigger and more powerful dilution refrigerators, 
contain ferromagnetic material that does not allow them to be placed close to the quantum processor, 
and cause losses that limit the readout fidelity of the measurement chain.
In recent years, with major advances in superconducting quantum circuits~\cite{Devoret_2013} and attempts to scale-up quantum processors, 
these shortcomings 
have called for chip-scale lossless nonreciprocal devices without ferromagnetic materials~\cite{kamal_noiseless_2011}. 

There have been numerous demonstrations of parametric nonreciprocal devices %
harnessing three-wave mixing in dc-SQUIDs or Josephson parametric converters (JPCs). 
A three-mode nonlinear superconducting circuit using a dc-SQUID and two linear LC-resonators %
was engineered to implement a reconfigurable frequency-converting 
three-port circulator or directional amplifier~%
\cite{ranzani_graph-based_2015,lecocq_nonreciprocal_2017}. 
Similar functionality was achieved using a single-stage JPC~%
\cite{sliwa_reconfigurable_2015}. 
Meanwhile, two coupled JPCs enabled implementation of frequency-preserving Josephson directional amplifiers~%
\cite{abdo_directional_2013,abdo_josephson_2014,abdo_multi-path_2018} 
and a gyrator~\cite{abdo_gyrator_2017}, 
which can easily be converted into a four-port circulator. 
A four-port on-chip superconducting circulator was also demonstrated using a combination of frequency conversion and delay~\cite{chapman_widely_2017}, 
which can be viewed as a ``synthetic rotation''~\cite{kerckhoff2015}. 

Another class of superconducting devices that exhibit directionality is travelling-wave parametric amplifiers (TWPAs)~\cite{ranzani_geometric_2014}. Near-quantum-limited TWPAs were realized employing nonlinearity of Josephson junctions~\cite{white2015,macklin_nearquantum-limited_2015} and kinetic  inductance~\cite{eom_wideband_2012,vissers_low-noise_2016}. The directionality of the TWPA gain arises from phase matching of the parametric interaction, since amplification occurs only when the signal is copropagating with the pump. Therefore, the signal travelling in the reverse direction stays unmodified: the reverse gain of an ideal TWPA is 0 dB. Another intrinsically directional amplifier utilizes a superconducting low-inductance undulatory galvanometer (SLUG, a version of a dc-SQUID)~\cite{Hover_superconducting_2012} in a finite voltage state. 
In contrast to TWPAs, SLUG microwave amplifiers feature reverse isolation comparable to commercial isolators~\cite{thorbeck_reverse_2017}.  

Similarly, nonreciprocal devices can be engineered using 
the nonlinearity of the optomechanical coupling~%
\cite{cavity_optomechanics_RMP} both in the optical and microwave domains. 
The first proposal of an optomechanical nonreciprocal device considered a travelling-wave geometry~%
\cite{hafezi_optomechanically_2012} 
and was implemented in the optical domain~%
\cite{shen_experimental_2016,ruesink2016}. 

A general approach to engineer nonreciprocal photon transport 
compatible with optomechanical systems was outlined by Metelmann and Clerk~%
\cite{metelmann_nonreciprocal_2015}. 
The framework was formulated in terms of interference 
between coherent and dissipative coupling channels 
linking two electromagnetic modes. 
Closely following this proposal, 
an optical isolator and directional amplifier was demonstrated~%
\cite{fang_generalized_2017}. 
An alternative scheme that does not employ direct coherent coupling between electromagnetic modes but only 
optomechanical interactions
was implemented with superconducting circuits resulting in electromechanical isolators and circulators~%
\cite{bernier_nonreciprocal_2017,peterson_demonstration_2017,barzanjeh_mechanical_2017}. 
Near-quantum-limited phase-preserving and phase-sensitive electromechanical amplifiers 
can be engineered in a similar way~\cite{malz_quantum-limited_2018}.

Here we explain the origin of nonreciprocity in electromechanical circuits 
by using the gyrator-based isolator as a conceptual starting point.
We then discuss the limitations of the current experimental implementations 
of electromechanical nonreciprocal devices 
as well as the future prospects.

\section{
The gyrator-based isolator
}
\label{sec:requirements}
%
%
%

\newcommand{\mec}{\mathrm{m}}
\newcommand{\inp}{\mathrm{in}}
\newcommand{\out}{\mathrm{out}}
\newcommand{\ext}{\mathrm{ex}}
\newcommand{\modea}{\hat a} 
\newcommand{\modeb}{\hat b}
\newcommand{\coh}{\mathrm{coh}}
\newcommand{\dis}{\mathrm{dis}}
Any linear two-port device can be described by its scattering matrix $S$,
linking the incoming modes
$\modea_{i,\inp}$
to the outgoing modes 
$\modea_{i,\out}$
by
\begin{equation}
  \begin{pmatrix}
    \modea_{1,\out}\\
    \modea_{2,\out}
  \end{pmatrix}
  = 
  \begin{pmatrix}
   S_{11} & S_{12} \\
   S_{21} & S_{22} 
  \end{pmatrix}
  \begin{pmatrix}
    \modea_{1,\inp}\\
    \modea_{2,\inp}
  \end{pmatrix}
  .
\end{equation}
A starting definition for nonreciprocity is that
a device is nonreciprocal when 
$S_{21}\neq S_{12}$.
Physically this corresponds to 
a modification of the scattering process
when input and output modes are interchanged.

\begin{figure}
  \includegraphics[width=0.48\textwidth]
  {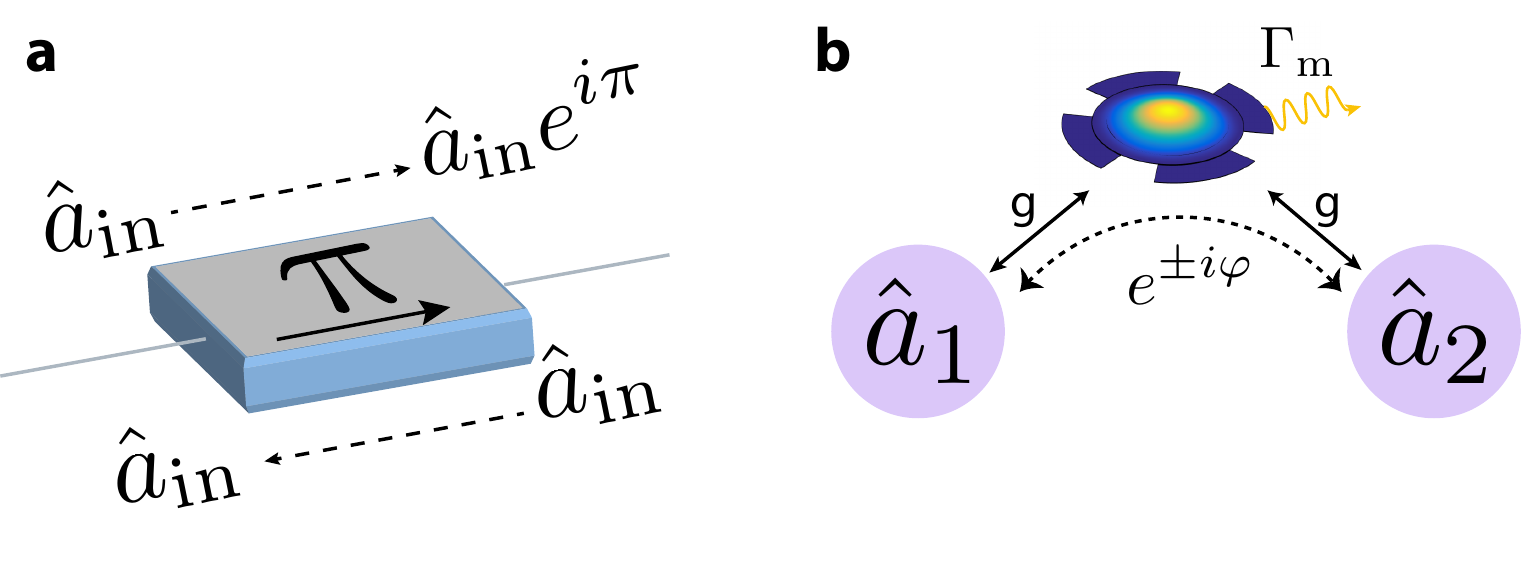}
  \caption{
  \textbf{Gyrator and optomechanical coupling.} 
  \textbf{a}.~ The gyrator is a canonical nonreciprocal component. 
  It is a two-port device, 
  which adds a $\pi$ phase shift when a wave is traveling one way
  but no phase shift in the reverse direction.
  \textbf{b}.~A simple multimode optomechanical system consists of two electromagnetic modes coupled to the same mechanical oscillator. 
  Due to the two microwave drive tones
  which linearize the optomechanical coupling, 
  the conversion from 
  $\modea_1$ to $\modea_2$ 
  formally imparts a nonreciprocal phase shift, similarly to the gyrator. 
  In the case of frequency conversion, 
  this phase shift between tones at different frequencies
  is not measureable, as it depends on the reference frame.
    \label{fig:figure1}
   }
\end{figure}

In order to develop an intuitive picture for nonreciprocity,
we introduce the gyrator as a canonical nonreciprocal element.
The gyrator is a 2-port device which provides a nonreciprocal phase shift~%
\cite{pozar_microwave_2011},
as illustrated in \cref{fig:figure1}a.
In one direction, it imparts a phase shift of
$\pi$, 
while in the other direction it leaves the signal unchanged. 
The scattering matrix of a lossless, matched gyrator is given by
$S_\mathrm{gyr} =  ( 
\begin{smallmatrix}
  0 & 1 \\ -1 & 0
\end{smallmatrix}
)$.
Recently, new implementations of microwave gyrators 
relying on the non-commutativity of frequency and time translations~%
\cite{rosenthal_breaking_2017}
have been demonstrated  with Josephson junctions~%
\cite{abdo_gyrator_2017,chapman_widely_2017}. %
The gyrator
in itself is not commonly used in technological applications,
but rather as a fundamental nonreciprocal building block
to construct other nonreciprocal devices.

An isolator is a useful nonreciprocal device
that can be assembled from a gyrator
and additional reciprocal elements~%
\cite{pozar_microwave_2011}.
A beam splitter can divide a signal in two;
one part goes through a gyrator 
while the other propagates with no phase shift.
Recombining the signals with a second beam splitter
results in a 4-port device that interferes the signal nonreciprocally,
as illustrated in \cref{fig:figure2}a.
For a signal injected in port 1,
the recombined signal after the second beam splitter 
interferes destructively in port 4,
but constructively in port 2.
In contrast, a signal injected from port 2,
reaches port 3 instead of port 1,
since one arm of the signal is in this case subjected to a $\pi$ shift.
Overall, the device is a four-port circulator 
that redirects each port to the next.
To obtain a two-port isolator, two of the four ports are terminated by
 matched loads to absorb the unwanted signal.
The scattering matrix for the remaining two ports 
is that of an ideal isolator,
$S_{\mathrm{is}}
=
(
\begin{smallmatrix}
  0 & 0 \\ 1 & 0
\end{smallmatrix}
).
$

The 
gyrator-based scheme
helps to summarize three sufficient ingredients 
to realize an isolator.
Firstly, an element \emph{breaks reciprocity}
by inducing a nonreciprocal phase shift.
Secondly, an additional path is introduced
for signals to \emph{interfere}  
such that the scattering matrix becomes asymmetric 
in the amplitude with 
$|S_{21}| \neq |S_{12}|$.
Finally, \emph{dissipation} is required for an isolator,
since its scattering matrix between the two ports 
is non-unitary and 
some signals must necessarily be redirected 
to an external degree of freedom.
The gyrator-based isolator
provides a framework to understand how nonreciprocity
arises in microwave optomechanical implementations.

\section{Nonreciprocity in optomechanical systems}
Microwave optomechanical schemes for nonreciprocity
rely on scattering between coupled modes~%
\cite{ranzani_geometric_2014,ranzani_graph-based_2015}.
As a first step towards  optomechanical isolators,
we introduce optomechanical frequency conversion~%
\cite{lecocq_mechanically_2016}
and how it relates to the gyrator.

The simplest optomechanical scheme to couple two electromagnetic modes
$\modea_1$ and $\modea_2$
involves coupling them both to the same mechanical oscillator
$\modeb$
(\cref{fig:figure1}b).
The optomechanical coupling terms
$\hbar g_{0,i} \,  \modea_i^\dagger \modea_i \, 
( \modeb + \modeb^\dagger )$
($i=1,2$),
where $g_{0,i}$
is the vacuum coupling rate of $\modea_i$ and $\modeb$,
can be linearized by two applied tones,
detuned by $\Delta_{i}$ with respect to each cavity resonance~%
\cite{cavity_optomechanics_RMP}.
In a frame rotating at the pump frequencies,
and keeping only the linear terms
and taking the rotating-wave approximation,
the effective Hamiltonian becomes~%
\cite{cavity_optomechanics_RMP}
\begin{multline}
  H
  =
  - \hbar \Delta_1 \modea_1^\dagger \modea_1
  - \hbar \Delta_2 \modea_2^\dagger \modea_2
  + \hbar \Omega_\mec \modeb^\dagger \modeb
  \\
  + \hbar g_1 \left( e^{i\phi_1} \modea_1 \modeb^\dagger 
  + e^{-i\phi_1} \modea_1^\dagger \modeb\right)
  + \hbar g_2 \left( e^{i\phi_2} \modea_2 \modeb^\dagger 
  + e^{-i\phi_2} \modea_2^\dagger \modeb\right)
  ,
  \label{eq:Hconv}
\end{multline}
where 
$\Omega_\mec$ is the mechanical frequency,
$g_i=g_{0,i} \sqrt{n_{c,i}}$
is the coupling rate enhanced by the photon number
$n_{c,i}$ due to the pump field,
and $\phi_i$ is the phase of each pump field.
For the resonant case
$\Delta_1 = \Delta_2 = -\Omega_\mec$,
the frequency conversion between the two modes through mechanical motion
is characterized by the scattering matrix elements
(at the center of the frequency conversion window)~%
\cite{safavi-naeini_proposal_2011}
\begin{equation}
  S_{21}
  =
  \frac{2\sqrt{\mathcal{C}_1 \mathcal{C}_2}}{1 + \mathcal{C}_1 + \mathcal{C}_2}
  e^{i(\phi_1 - \phi_2)}
  \;\; \text{and} \;\;
  S_{12}
  =
  S_{21}^*
  ,
  \label{eq:FreqConvS}
\end{equation}
where
$\kappa_i$ is the energy decay rate of mode $\modea_i$,
and 
$\mathcal{C}_i = 4 g_i^2 / (\kappa_i \Gamma_\mec)$
is the cooperativity with 
$\Gamma_\mec$
the energy decay rate of the mechanical oscillator
(for simplicity, the cavities are assumed to be overcoupled).

\begin{figure}[t]
  \includegraphics
  [width=\columnwidth]
  {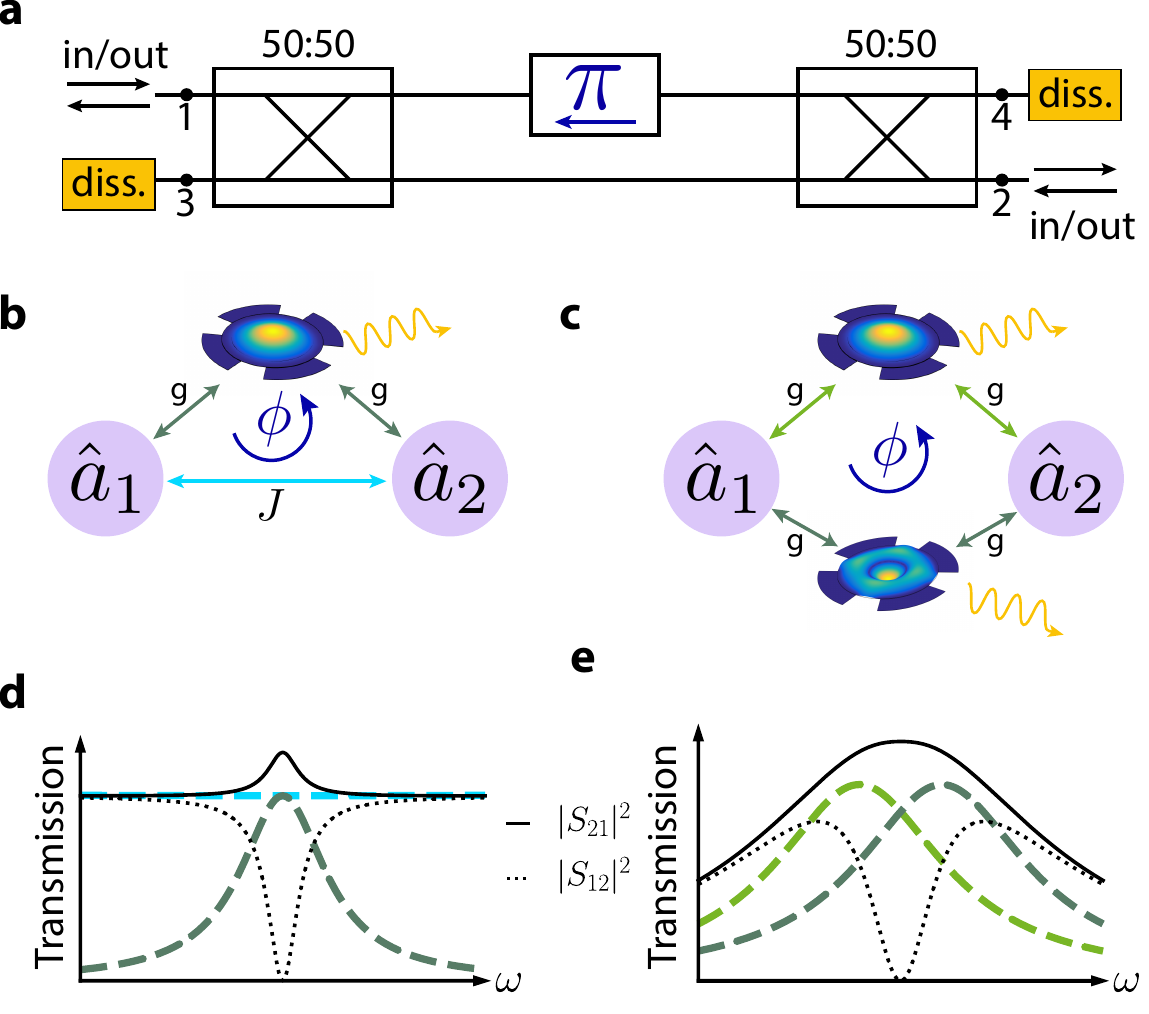}
  \caption{
  \textbf{
  Gyrator-based isolator compared to optomechanical multimode schemes.
  }
  \textbf{a}.~An isolator can be built by combining the gyrator 
  with other reciprocal elements. 
  By combining the gyrator with two beam splitters and a transmission line,
  a four-port circulator is realized.
  Dissipation (provided by line terminations) 
  eliminates the unwanted ports.
  \textbf{b}.~The three-mode optomechanical isolator 
  in which two electromagnetic modes have a direct coherent $J$-coupling,
  and interact through a shared mechanical mode. 
    \textbf{c}.~The four-mode optomechanical isolator,
    in which two electromagnetic modes interact through 
    to different mechanical modes.
    \textbf{d}.~In the scheme (b), frequency conversion through
    the $J$-coupling (light blue) and through the mechanical modes (green)
    interfere differently in the forward and backward direction when combined.
    \textbf{e}.~In the scheme (c), the frequency conversion
    through each mechanical mode (green curves)
    are offset in frequency
    to interfere in a nonreciprocal manner.
    \label{fig:figure2}
   }
\end{figure} 

While \cref{eq:FreqConvS} apparently fulfills the condition
$S_{21} \neq S_{12}$ for nonreciprocity,
the situation is more subtle due to the fact that the two modes
are in fact at different frequencies in the laboratory frame.
The time-dependence for the two modes can be written explicitly
to understand where the issue lies.
The first incoming mode
$\modea_{1,\inp} (t) = e^{-\omega_1 t} A_1$
results in 
$\modea_{2,\out} (t) = S_{21} e^{-i\omega_2 t} A_1$
and reciprocally the other mode
$\modea_{2,\inp} (t) = e^{-i\omega_2 t} A_2$
results in 
$\modea_{1,\out} (t) = S_{12} e^{-i\omega_1 t} A_2$,
where 
$\omega_i$ are the mode angular frequencies
and $A_i$ are constant amplitudes.
If a new origin of time $t_0$
is chosen,
the fields transform as
$\modea_{i}' 
(t) 
= e^{-i\omega_i t_0} a_{i}
(t)$,
equivalent to a frequency-dependent phase shift for each mode.
In this new reference frame, the scattering matrix transforms as
$S_{21}' = 
S_{21} e^{-i(\omega_2 - \omega_1)t_0}$
and
$S_{12}' = 
S_{12} e^{+i(\omega_2 - \omega_1)t_0}$.
For different frequencies $\omega_1 \neq \omega_2$,
there always exists a $t_0$
for which the phases of $S_{21}$ and $S_{12}$ are the same.
A nonreciprocal phase shift is therefore unphysical 
for frequency conversion,
as the phase depends on the chosen reference frame.
For this reason, Ranzani and Aumentado~%
\cite{ranzani_graph-based_2015}
pose the stricter requirement
$|S_{12}| \neq |S_{21}|$ 
for nonreciprocity in coupled-modes systems.
The pump tones that linearize the optomechanical coupling
break reciprocity,
as they are held fixed when input and output are interchanged
and impose a fixed phase $\phi_1- \phi_2$ for the coupling.
Nevertheless, there is always a frame with a different origin of time
in which the two pumps have the same phase and 
$\phi'_1 - \phi'_2 = 0$.
In that frame, the symmetry between the two ports 
is apparently restored.

While a change of the origin of time turns
a phase-nonreciprocal system reciprocal for coupled-mode systems,
it simultaneously turns
reciprocal systems phase-nonreciprocal.
The isolator in \cref{fig:figure1}B provides an example
(%
adding frequency conversion as a thought experiment%
).
With 
$(\omega_2 - \omega_1)t_0 = \pi/2$,
the gyrator scattering matrix transforms from
$S_\mathrm{gyr} =
(
\begin{smallmatrix}
  0 & 1 \\ -1 & 0
\end{smallmatrix}
)
$
to 
$S'_\mathrm{gyr} =
e^{i\pi/2}
(
\begin{smallmatrix}
  0 & 1 \\ 1 & 0
\end{smallmatrix}
)
$
while the transmission line scattering matrix transforms from
$S_\mathrm{tl} =
(
\begin{smallmatrix}
  0 & 1 \\ 1 & 0
\end{smallmatrix}
)
$
to
$S'_\mathrm{tl} =
e^{i\pi/2}
(
\begin{smallmatrix}
  0 & 1 \\ -1 & 0
\end{smallmatrix}
)
$.
In effect, gyrators and transmission lines are mapped to each other.
Importantly,
the combination of a gyrator and a transmission line
is preserved.
One interpretation is that the frame-dependent
nonreciprocal phase acts as a gauge symmetry,
that can realize the Aharonov-Bohm effect
when a loop is created~%
\cite{fang_photonic_2012,fang_experimental_2013,%
tzuang_non-reciprocal_2014}.
To build an optomechanical isolator, one method
is to realize two paths between $\modea_1$ and $\modea_2$,
one similar to a gyrator and the other to a transmission line,
which can be done following one of two proposed schemes.

The first scheme, shown in \cref{fig:figure2}b,
combines an optomechanical link between two electromagnetic modes
$\modea_1$ and $\modea_2$
and a direct coherent coupling of strength $J$.
The interaction term of the latter, given by
$H_\coh =
\hbar J  ( 
e^{i\theta} \modea_1 \modea_2^\dagger + e^{-i\theta} \modea_1^\dagger \modea_2
)$
induces by itself conversion between the two modes as
\begin{equation}
  S_{21}
  =
  \frac{2 \sqrt{\mathcal{C}_\coh}}{1 + \mathcal{C}_\coh}
  i e^{i\theta}
  \quad \text{and} \quad
  S_{12}
  =
  \frac{2 \sqrt{\mathcal{C}_\coh}}{1 + \mathcal{C}_\coh}
  i e^{-i\theta}
\end{equation}
where
$\mathcal{C}_\coh = 4 J^2 /(\kappa_1 \kappa_2)$.
Compared to \cref{eq:FreqConvS},
there is an intrinsic reciprocal phase shift of 
$i = e^{i\pi/2}$.
By choosing the phases 
$\phi_1=\phi_2$ and $\theta = \pi/2$,
the optomechanical link realizes a reciprocal transmission line
while the coherent link breaks the symmetry
and realizes a gyrator.
The combination,
with matching coupling rates,
functions as an isolator between the modes 
$\modea_1$ and $\modea_2$.
One cannot tell 
which 
of the optomechanical or direct link
breaks reciprocity,
as it depends on the chosen frame.
Nonetheless, there is a gauge-invariant phase
$\phi = \phi_1 - \phi_2 + \theta$
that globally characterizes the broken symmetry
and can be seen as a synthetic magnetic flux
\cite{peano_topological_2015,fang_generalized_2017}.

The second scheme to realize isolation, shown in \cref{fig:figure2}c,  
uses two optomechanical conversion links 
with two different mechanical modes.
From \cref{eq:FreqConvS}, 
one cannot
realize the scattering matrices of a gyrator and a transmission line
with the same reciprocal phase factor.
Dissipation of the mechanical modes is the key 
to tune the overall reciprocal phase factor of conversion.
By detuning the two mechanical modes in frequency,
the mechanical susceptibilities induce different phases.
%
Advantageously over the first scheme,
no direct coherent coupling must be engineered and
the modes 
$\modea_1$ and $\modea_2$
do not need to have the same frequency.

The first scheme (\cref{fig:figure2}b)
derives from a proposal by Metelmann and Clerk~%
\cite{metelmann_nonreciprocal_2015},
who describe the optomechanical link
as an effective dissipative interaction between
the modes
$\modea_1$ and $\modea_2$.
This coupling is equivalent to
a non-Hermitian Hamiltonian term
$
H_{\dis}
=
- i \hbar \Gamma_\dis
(
  e^{i(\phi_1 - \phi_2)} \modea_1 \modea_2^\dagger
  +
  e^{-i(\phi_1 - \phi_2)} \modea_1^\dagger \modea_2
)
$
with 
$\Gamma_\dis = 2 g_1 g_2 / \Gamma_\mec$.
With suitable parameters, the total interaction
$H_\coh + H_\dis$
can be made unidirectional, 
with for instance a term proportional to 
$\modea_1^\dagger \modea_2$
but not
$\modea_1 \modea_2^\dagger$.
We note that
while only the case of resonant modes is considered here,
a direct coherent $J$-coupling can also be realized between
modes at different frequencies using parametric interactions.
The second scheme 
(\cref{fig:figure2}c)
can also be understood in that framework,
as each detuned mechanical conversion link is equivalent
to an effective interaction between the modes
$\modea_1$ and $\modea_2$
with both a coherent and a dissipative component.
The total effective interaction can be made unidirectional
when the total direct coherent interaction matches
the total dissipative interaction
and they interfere.

Both methods to achieve nonreciprocal isolation can be decomposed 
in terms of the three ingredients identified
in the gyrator-based isolator.
At the heart is the breaking of reciprocity
that occurs through the time-dependent drives applied to the system
that impart a complex phase to the interaction.
Then a loop constituted by two arms allow
the unwanted signal to be canceled in the backward direction 
through destructive interference
while preserving it in the forward direction.
The mechanical dissipative baths are used to eliminate
the backward signal.
As a consequence,  
the nonreciprocal bandwidth 
(where $S_{21}$ and $S_{12}$ differ)
is limited by the mechanical dissipation rates,
as illustrated in \cref{fig:figure2}d and e.
Dissipation plays a double role here, 
since it also gives
the different reciprocal phase shift in each arms
necessary to implement 
both a gyrator and a transmission line.


\section{Experimental realizations}
Microwave optomechanics is 
uniquely suited to implement the mode structure presented in 
\cref{fig:figure2}b,c 
and
achieve the nonreciprocal frequency conversion described above.
The electromagnetic (EM) modes are here modes
of a superconductive LC resonator 
and the mechanical modes are 
modes of the vibrating top plate of a vacuum-gap capacitor. 

\begin{figure}[t]
  \includegraphics
  [width=\columnwidth]
  {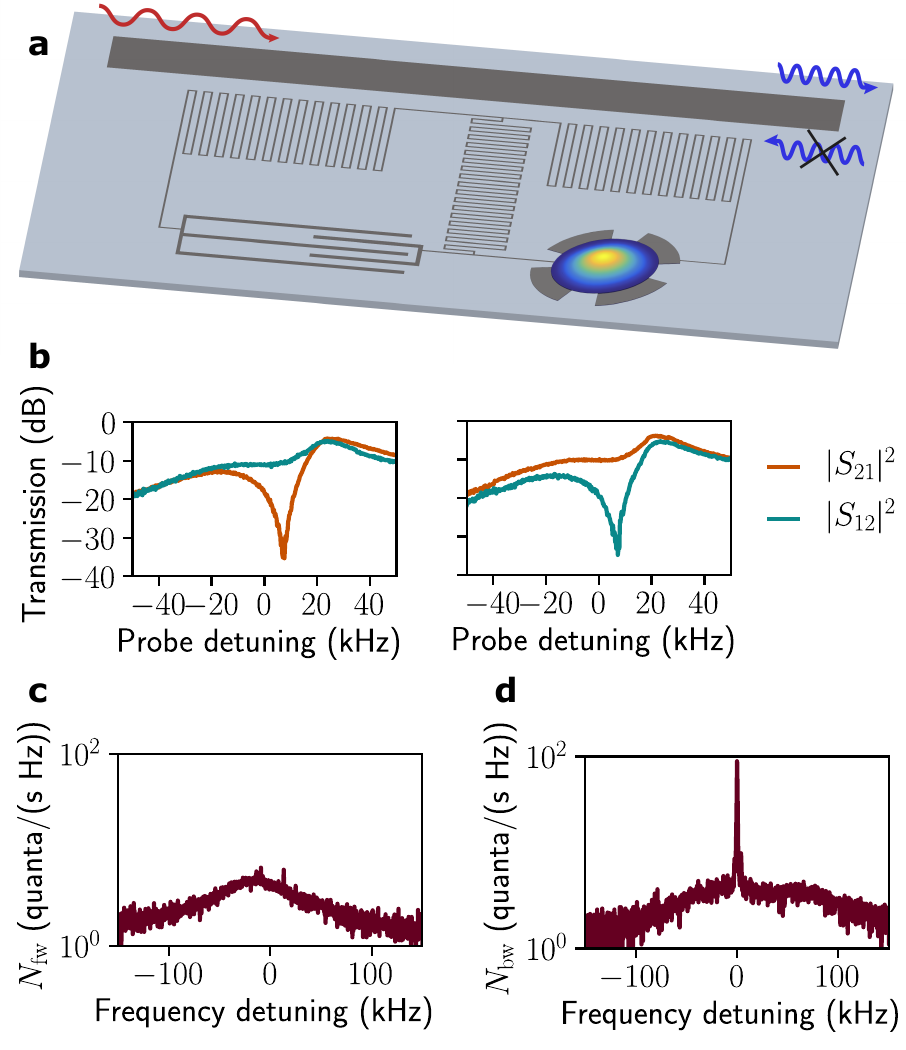}
  \caption{
  \textbf{Experimental realization of nonreciprocal frequency conversion 
  in a microwave optomechanical circuit.}
  (figure adapted from Bernier {et al.}~\cite{bernier_nonreciprocal_2017})
  \textbf{a}.~ In order the realize the mode structure of 
  \cref{fig:figure2}{c}, 
  a multimode electromechanical circuit can be constructed. 
  In the example shown, the circuit supports two modes, 
  each coupled to the motion of the top plate of a vacuum-gap capacitor 
  that supports two vibrational modes.
  \textbf{b}.~Transmission linear response of the frequency
  conversion between the modes $\modea_1$ and $\modea_2$, in both direction,
  for two phases $\phi$ for which the circuit acts as an isolator
  in one direction and the other.
  \textbf{c}.~Noise emission $N_\mathrm{fw}$
  in the forward direction of the isolator, 
  corresponding to the added noise to the signal.
  \textbf{d}.~Noise emission $N_\mathrm{bw}$
  in the backward direction, displaying large power
  in the isolation bandwidth, 
  as the mechanical bath noise is directly coupled to the input port.
      \label{fig:figure3}
   }
 \end{figure}

It turns out that the first, seemingly simplest, scheme 
shown in \cref{fig:figure2}c, 
in which two EM modes are coupled to one mechanical element 
as well as to each other through a direct coherent coupling, 
poses engineering challenges.
The two EM modes must have the same resonance frequencies
(within their linewidth)
and the fixed direct $J$-coupling must be strong enough to achieve
$\coop_\coh\approx1$.
Despite these difficulties, 
an analogous scheme has been implemented in the optical domain, 
in which 
two photonic-crystal modes are each coupled 
to a phononic-crystal mode 
and two direct coherent optical-optical and mechanical-mechanical link 
are realized through waveguide buses~%
\cite{fang_generalized_2017}.
The scheme is equivalent to the mode structure in 
\cref{fig:figure2}b
as the effective interaction through the two mechanical modes 
is similar to that through a single one.

By contrast, 
the simpler plaquette of the second scheme 
in \cref{fig:figure2}c,
that requires only optomechanical interactions and no frequency tuning,
has been realized in microwave optomechanical circuits~%
\cite{bernier_nonreciprocal_2017,peterson_demonstration_2017,%
barzanjeh_mechanical_2017}.
Two microwave modes 
of a superconducting circuit are
both coupled to two vibrational modes of the top plate of a capacitor.
An example circuit is shown in \cref{fig:figure3}a~%
\cite{bernier_nonreciprocal_2017}.
These systems are inherently frequency converting as 
the two microwave modes have different resonance frequencies.
The optomechanical couplings are established through
four phase-locked microwave pump tones
slightly detuned from the lower motional sidebands of each mode.
Two frequency conversion windows open, one through each mechanical mode,
and interfere with each other.
By varying the relative phases of each pump tone,
the gauge-independent phase $\phi$ can be tuned,
realizing the isolator scheme of
\cref{fig:figure2}c,e.
An example of the measured nonreciprocal transmission is shown 
in \cref{fig:figure3}b
for two values of $\phi$ for which the system isolates
in each direction.
There is no limitation on the achievable depth of isolation in these systems. 
However, since the scheme relies on the mechanical dissipation
to absorb the unwanted backward signal,
the nonreciprocal bandwidth is limited the bare energy decay rate 
of the mechanical modes, typically in the 10 - 100~Hz range,
although it can be effectively increased by external damping.



An important aspect of the nonreciprocal devices based on the optomechanical interaction is their noise performance.
It is apparent from the gyrator scheme presented 
in \cref{sec:requirements} and shown 
in \cref{fig:figure2}a 
that the dissipative elements terminating the unwanted ports modes 
in turn emit back their respective thermal noise into the system. 
In the forward direction 
(\cref{fig:figure3}c),
the emitted noise, corresponding to the added noise to the signal, 
is spread over a wide bandwidth 
and can be made quantum-limited in the high-cooperativity regime.
In the backward direction however
(\cref{fig:figure3}d),
noise commensurate with the high thermal occupancy of the
mechanical baths is directly emitted back from the input port,
within the isolation bandwidth.
External cooling of the mechanical modes is therefore required
for the optomechanical isolator to protect
sensitive quantum devices from back-propagating noise.


\section{Conclusion and outlook}
Microwave optomechanical circuits form a new platform
to design nonreciprocal devices that could be integrated
with superconducting quantum circuits.
Current realizations for isolators suffer from a narrow bandwidth,
limited by the mechanical dissipation rates,
as well as from high back-propagating noise.
External cooling of the mechanical oscillators
can mitigate both issues.
Microwave circulators realized with the same technology
\cite{barzanjeh_mechanical_2017}
provide a better solution
leading to quantum-limited noise 
and a nonreciprocal bandwidth only limited by 
the dissipation rates of the cavities~%
\cite{bernier_nonreciprocal_2017}.
Furthermore, directional amplification can be implemented
with the same structures as isolators,
both phase-preserving and quadrature dependent,
with prospects to reach the quantum limit in terms of added noise~%
\cite{malz_quantum-limited_2018}.
Optomechanical nonreciprocal devices have intrinsic limits to their bandwidth,
but their frequencies can be made adjustable~%
\cite{andrews_quantumenabled_2015}
and certain applications might benefit
from the versatility of such tunable narrow-band nonreciprocal devices.

\textit{Comment} This manuscript was submitted as a contribution to a special issue titled 
``Magnet-less Nonreciprocity in Electromagnetics" to appear in IEEE Antennas and Wireless Propagation Letters.

\begin{acknowledgments}
This work was supported by
the SNF, 
the NCCR QSIT,  
and the EU H2020 programme under grant agreement No 732894
(FET Proactive HOT). 
T.J.K. acknowledges financial support from an ERC AdG (QuREM).
\end{acknowledgments}

\bibliography{bibliography}

\begin{thebibliography}{10}
\providecommand{\url}[1]{#1}
\csname url@samestyle\endcsname
\providecommand{\newblock}{\relax}
\providecommand{\bibinfo}[2]{#2}
\providecommand{\BIBentrySTDinterwordspacing}{\spaceskip=0pt\relax}
\providecommand{\BIBentryALTinterwordstretchfactor}{4}
\providecommand{\BIBentryALTinterwordspacing}{\spaceskip=\fontdimen2\font plus
\BIBentryALTinterwordstretchfactor\fontdimen3\font minus
  \fontdimen4\font\relax}
\providecommand{\BIBforeignlanguage}[2]{{%
\expandafter\ifx\csname l@#1\endcsname\relax
\typeout{** WARNING: IEEEtran.bst: No hyphenation pattern has been}%
\typeout{** loaded for the language `#1'. Using the pattern for}%
\typeout{** the default language instead.}%
\else
\language=\csname l@#1\endcsname
\fi
#2}}
\providecommand{\BIBdecl}{\relax}
\BIBdecl

\bibitem{Devoret_2013}
M.~H. Devoret and R.~J. Schoelkopf, ``Superconducting circuits for quantum
  information: An outlook,'' \emph{Science}, vol. 339, p. 1169, 2013.

\bibitem{kamal_noiseless_2011}
\BIBentryALTinterwordspacing
A.~Kamal, J.~Clarke, and M.~H. Devoret, ``\BIBforeignlanguage{en}{Noiseless
  non-reciprocity in a parametric active device},''
  \emph{\BIBforeignlanguage{en}{Nature Physics}}, vol.~7, no.~4, pp. 311--315,
  Apr. 2011. [Online]. Available:
  \url{http://www.nature.com/nphys/journal/v7/n4/abs/nphys1893.html}
\BIBentrySTDinterwordspacing

\bibitem{ranzani_graph-based_2015}
\BIBentryALTinterwordspacing
L.~Ranzani and J.~Aumentado, ``\BIBforeignlanguage{en}{Graph-based analysis of
  nonreciprocity in coupled-mode systems},'' \emph{\BIBforeignlanguage{en}{New
  Journal of Physics}}, vol.~17, no.~2, p. 023024, 2015. [Online]. Available:
  \url{http://stacks.iop.org/1367-2630/17/i=2/a=023024}
\BIBentrySTDinterwordspacing

\bibitem{lecocq_nonreciprocal_2017}
\BIBentryALTinterwordspacing
F.~Lecocq, L.~Ranzani, G.~A. Peterson, K.~Cicak, R.~W. Simmonds, J.~D. Teufel,
  and J.~Aumentado, ``Nonreciprocal microwave signal processing with a
  field-programmable josephson amplifier,'' \emph{Phys. Rev. Applied}, vol.~7,
  p. 024028, Feb 2017. [Online]. Available:
  \url{https://link.aps.org/doi/10.1103/PhysRevApplied.7.024028}
\BIBentrySTDinterwordspacing

\bibitem{sliwa_reconfigurable_2015}
\BIBentryALTinterwordspacing
K.~Sliwa, M.~Hatridge, A.~Narla, S.~Shankar, L.~Frunzio, R.~Schoelkopf, and
  M.~Devoret, ``Reconfigurable {Josephson} {Circulator}/{Directional}
  {Amplifier},'' \emph{Physical Review X}, vol.~5, no.~4, p. 041020, Nov. 2015.
  [Online]. Available: \url{http://link.aps.org/doi/10.1103/PhysRevX.5.041020}
\BIBentrySTDinterwordspacing

\bibitem{abdo_directional_2013}
\BIBentryALTinterwordspacing
B.~Abdo, K.~Sliwa, L.~Frunzio, and M.~Devoret, ``Directional {Amplification}
  with a {Josephson} {Circuit},'' \emph{Physical Review X}, vol.~3, no.~3, p.
  031001, Jul. 2013. [Online]. Available:
  \url{http://link.aps.org/doi/10.1103/PhysRevX.3.031001}
\BIBentrySTDinterwordspacing

\bibitem{abdo_josephson_2014}
\BIBentryALTinterwordspacing
B.~Abdo, K.~Sliwa, S.~Shankar, M.~Hatridge, L.~Frunzio, R.~Schoelkopf, and
  M.~Devoret, ``Josephson {Directional} {Amplifier} for {Quantum} {Measurement}
  of {Superconducting} {Circuits},'' \emph{Physical Review Letters}, vol. 112,
  no.~16, p. 167701, Apr. 2014. [Online]. Available:
  \url{http://link.aps.org/doi/10.1103/PhysRevLett.112.167701}
\BIBentrySTDinterwordspacing

\bibitem{abdo_multi-path_2018}
\BIBentryALTinterwordspacing
B.~Abdo, N.~T. Bronn, O.~Jinka, S.~Olivadese, M.~Brink, and J.~M. Chow,
  ``Multi-path interferometric josephson directional amplifier for qubit
  readout,'' \emph{Quantum Science and Technology}, vol.~3, no.~2, p. 024003,
  2018. [Online]. Available:
  \url{http://stacks.iop.org/2058-9565/3/i=2/a=024003}
\BIBentrySTDinterwordspacing

\bibitem{abdo_gyrator_2017}
B.~Abdo, M.~Brink, and J.~M. Chow, ``Gyrator {{Operation Using Josephson
  Mixers}},'' \emph{Physical Review Applied}, vol.~8, no.~3, p. 034009, Sep.
  2017.

\bibitem{chapman_widely_2017}
B.~J. Chapman, E.~I. Rosenthal, J.~Kerckhoff, B.~A. Moores, L.~R. Vale,
  J.~A.~B. Mates, G.~C. Hilton, K.~Lalumi{\`e}re, A.~Blais, and K.~W. Lehnert,
  ``Widely {{Tunable On}}-{{Chip Microwave Circulator}} for {{Superconducting
  Quantum Circuits}},'' \emph{Physical Review X}, vol.~7, no.~4, p. 041043,
  Nov. 2017.

\bibitem{kerckhoff2015}
J.~Kerckhoff, K.~Lalumi{\`e}re, B.~J. Chapman, A.~Blais, and K.~W. Lehnert,
  ``On-{{Chip Superconducting Microwave Circulator}} from {{Synthetic
  Rotation}},'' \emph{Physical Review Applied}, vol.~4, no.~3, p. 034002, Sep.
  2015.

\bibitem{ranzani_geometric_2014}
\BIBentryALTinterwordspacing
L.~Ranzani and J.~Aumentado, ``\BIBforeignlanguage{en}{A geometric description
  of nonreciprocity in coupled two-mode systems},''
  \emph{\BIBforeignlanguage{en}{New Journal of Physics}}, vol.~16, no.~10, p.
  103027, 2014. [Online]. Available:
  \url{http://stacks.iop.org/1367-2630/16/i=10/a=103027}
\BIBentrySTDinterwordspacing

\bibitem{white2015}
T.~C. White, J.~Y. Mutus, I.-C. Hoi, R.~Barends, B.~Campbell, Y.~Chen, Z.~Chen,
  B.~Chiaro, A.~Dunsworth, E.~Jeffrey, J.~Kelly, A.~Megrant, C.~Neill, P.~J.~J.
  O'Malley, P.~Roushan, D.~Sank, A.~Vainsencher, J.~Wenner, S.~Chaudhuri,
  J.~Gao, and J.~M. Martinis, ``Traveling wave parametric amplifier with
  {{Josephson}} junctions using minimal resonator phase matching,''
  \emph{Applied Physics Letters}, vol. 106, no.~24, p. 242601, Jun. 2015.

\bibitem{macklin_nearquantum-limited_2015}
\BIBentryALTinterwordspacing
C.~Macklin, K.~O'Brien, D.~Hover, M.~E. Schwartz, V.~Bolkhovsky, X.~Zhang,
  W.~D. Oliver, and I.~Siddiqi, ``\BIBforeignlanguage{en}{A
  near--quantum-limited {Josephson} traveling-wave parametric amplifier},''
  \emph{\BIBforeignlanguage{en}{Science}}, vol. 350, no. 6258, pp. 307--310,
  Oct. 2015. [Online]. Available:
  \url{http://science.sciencemag.org/content/350/6258/307}
\BIBentrySTDinterwordspacing

\bibitem{eom_wideband_2012}
\BIBentryALTinterwordspacing
B.~Ho~Eom, P.~K. Day, H.~G. LeDuc, and J.~Zmuidzinas, ``A wideband, low-noise
  superconducting amplifier with high dynamic range,'' \emph{Nature Physics},
  vol.~8, pp. 623--627, 07 2012. [Online]. Available:
  \url{http://dx.doi.org/10.1038/nphys2356}
\BIBentrySTDinterwordspacing

\bibitem{vissers_low-noise_2016}
\BIBentryALTinterwordspacing
M.~R. Vissers, R.~P. Erickson, H.-S. Ku, L.~Vale, X.~Wu, G.~C. Hilton, and
  D.~P. Pappas, ``Low-noise kinetic inductance traveling-wave amplifier using
  three-wave mixing,'' \emph{Applied Physics Letters}, vol. 108, no.~1, p.
  012601, 2016. [Online]. Available: \url{https://doi.org/10.1063/1.4937922}
\BIBentrySTDinterwordspacing

\bibitem{Hover_superconducting_2012}
\BIBentryALTinterwordspacing
D.~Hover, Y.-F. Chen, G.~J. Ribeill, S.~Zhu, S.~Sendelbach, and R.~McDermott,
  ``Superconducting low-inductance undulatory galvanometer microwave
  amplifier,'' \emph{Applied Physics Letters}, vol. 100, no.~6, p. 063503,
  2012. [Online]. Available: \url{https://doi.org/10.1063/1.3682309}
\BIBentrySTDinterwordspacing

\bibitem{thorbeck_reverse_2017}
\BIBentryALTinterwordspacing
T.~Thorbeck, S.~Zhu, E.~Leonard, R.~Barends, J.~Kelly, J.~M. Martinis, and
  R.~McDermott, ``Reverse isolation and backaction of the slug microwave
  amplifier,'' \emph{Phys. Rev. Applied}, vol.~8, p. 054007, Nov 2017.
  [Online]. Available:
  \url{https://link.aps.org/doi/10.1103/PhysRevApplied.8.054007}
\BIBentrySTDinterwordspacing

\bibitem{cavity_optomechanics_RMP}
\BIBentryALTinterwordspacing
M.~Aspelmeyer, T.~J. Kippenberg, and F.~Marquardt, ``Cavity optomechanics,''
  \emph{Rev. Mod. Phys.}, vol.~86, pp. 1391--1452, Dec 2014. [Online].
  Available: \url{http://link.aps.org/doi/10.1103/RevModPhys.86.1391}
\BIBentrySTDinterwordspacing

\bibitem{hafezi_optomechanically_2012}
\BIBentryALTinterwordspacing
M.~Hafezi and P.~Rabl, ``\BIBforeignlanguage{EN}{Optomechanically induced
  non-reciprocity in microring resonators},''
  \emph{\BIBforeignlanguage{EN}{Optics Express}}, vol.~20, no.~7, pp.
  7672--7684, Mar. 2012. [Online]. Available:
  \url{http://www.osapublishing.org/abstract.cfm?uri=oe-20-7-7672}
\BIBentrySTDinterwordspacing

\bibitem{shen_experimental_2016}
\BIBentryALTinterwordspacing
Z.~Shen, Y.-L. Zhang, Y.~Chen, C.-L. Zou, Y.-F. Xiao, X.-B. Zou, F.-W. Sun,
  G.-C. Guo, and C.-H. Dong, ``\BIBforeignlanguage{en}{Experimental realization
  of optomechanically induced non-reciprocity},''
  \emph{\BIBforeignlanguage{en}{Nature Photonics}}, vol.~10, no.~10, pp.
  657--661, Oct. 2016. [Online]. Available:
  \url{http://www.nature.com/nphoton/journal/v10/n10/full/nphoton.2016.161.html}
\BIBentrySTDinterwordspacing

\bibitem{ruesink2016}
F.~Ruesink, M.-A. Miri, A.~Al{\`u}, and E.~Verhagen, ``Nonreciprocity and
  magnetic-free isolation based on optomechanical interactions,'' \emph{Nature
  Communications}, vol.~7, p. 13662, Nov. 2016.

\bibitem{metelmann_nonreciprocal_2015}
\BIBentryALTinterwordspacing
A.~Metelmann and A.~Clerk, ``Nonreciprocal {Photon} {Transmission} and
  {Amplification} via {Reservoir} {Engineering},'' \emph{Physical Review X},
  vol.~5, no.~2, p. 021025, Jun. 2015. [Online]. Available:
  \url{http://link.aps.org/doi/10.1103/PhysRevX.5.021025}
\BIBentrySTDinterwordspacing

\bibitem{fang_generalized_2017}
\BIBentryALTinterwordspacing
K.~Fang, J.~Luo, A.~Metelmann, M.~H. Matheny, F.~Marquardt, A.~A. Clerk, and
  O.~Painter, ``Generalized non-reciprocity in an optomechanical circuit via
  synthetic magnetism and reservoir engineering,'' \emph{Nat Phys}, vol.~13,
  no.~5, pp. 465--471, 05 2017. [Online]. Available:
  \url{http://dx.doi.org/10.1038/nphys4009}
\BIBentrySTDinterwordspacing

\bibitem{bernier_nonreciprocal_2017}
\BIBentryALTinterwordspacing
N.~R. Bernier, L.~D. T{\'o}th, A.~Koottandavida, M.~A. Ioannou, D.~Malz,
  A.~Nunnenkamp, A.~K. Feofanov, and T.~J. Kippenberg,
  ``\BIBforeignlanguage{En}{Nonreciprocal reconfigurable microwave
  optomechanical circuit},'' \emph{\BIBforeignlanguage{En}{Nature
  Communications}}, vol.~8, no.~1, p. 604, Sep. 2017. [Online]. Available:
  \url{https://www.nature.com/articles/s41467-017-00447-1}
\BIBentrySTDinterwordspacing

\bibitem{peterson_demonstration_2017}
\BIBentryALTinterwordspacing
G.~Peterson, F.~Lecocq, K.~Cicak, R.~Simmonds, J.~Aumentado, and J.~Teufel,
  ``Demonstration of {Efficient} {Nonreciprocity} in a {Microwave}
  {Optomechanical} {Circuit},'' \emph{Physical Review X}, vol.~7, no.~3, p.
  031001, Jul. 2017. [Online]. Available:
  \url{https://link.aps.org/doi/10.1103/PhysRevX.7.031001}
\BIBentrySTDinterwordspacing

\bibitem{barzanjeh_mechanical_2017}
\BIBentryALTinterwordspacing
S.~Barzanjeh, M.~Wulf, M.~Peruzzo, M.~Kalaee, P.~B. Dieterle, O.~Painter, and
  J.~M. Fink, ``\BIBforeignlanguage{En}{Mechanical on-chip microwave
  circulator},'' \emph{\BIBforeignlanguage{En}{Nature Communications}}, vol.~8,
  no.~1, p. 953, Oct. 2017. [Online]. Available:
  \url{https://www.nature.com/articles/s41467-017-01304-x}
\BIBentrySTDinterwordspacing

\bibitem{malz_quantum-limited_2018}
\BIBentryALTinterwordspacing
D.~Malz, L.~D. T\'oth, N.~R. Bernier, A.~K. Feofanov, T.~J. Kippenberg, and
  A.~Nunnenkamp, ``Quantum-limited directional amplifiers with optomechanics,''
  \emph{Phys. Rev. Lett.}, vol. 120, p. 023601, Jan 2018. [Online]. Available:
  \url{https://link.aps.org/doi/10.1103/PhysRevLett.120.023601}
\BIBentrySTDinterwordspacing

\bibitem{pozar_microwave_2011}
D.~M. Pozar, \emph{Microwave Engineering}, 4th~ed.\hskip 1em plus 0.5em minus
  0.4em\relax Wiley, 2011.

\bibitem{rosenthal_breaking_2017}
E.~I. Rosenthal, B.~J. Chapman, A.~P. Higginbotham, J.~Kerckhoff, and K.~W.
  Lehnert, ``Breaking {{Lorentz Reciprocity}} with {{Frequency Conversion}} and
  {{Delay}},'' \emph{Physical Review Letters}, vol. 119, no.~14, p. 147703,
  Oct. 2017.

\bibitem{lecocq_mechanically_2016}
\BIBentryALTinterwordspacing
F.~Lecocq, J.~Clark, R.~Simmonds, J.~Aumentado, and J.~Teufel, ``Mechanically
  {Mediated} {Microwave} {Frequency} {Conversion} in the {Quantum} {Regime},''
  \emph{Physical Review Letters}, vol. 116, no.~4, p. 043601, Jan. 2016.
  [Online]. Available:
  \url{http://link.aps.org/doi/10.1103/PhysRevLett.116.043601}
\BIBentrySTDinterwordspacing

\bibitem{safavi-naeini_proposal_2011}
A.~H. Safavi-Naeini and O.~Painter, ``\BIBforeignlanguage{en}{Proposal for an
  optomechanical traveling wave phonon\textendash{}photon translator},''
  \emph{\BIBforeignlanguage{en}{New Journal of Physics}}, vol.~13, no.~1, p.
  013017, 2011.

\bibitem{fang_photonic_2012}
K.~Fang, Z.~Yu, and S.~Fan, ``Photonic {{Aharonov}}-{{Bohm Effect Based}} on
  {{Dynamic Modulation}},'' \emph{Physical Review Letters}, vol. 108, no.~15,
  p. 153901, Apr. 2012.

\bibitem{fang_experimental_2013}
------, ``Experimental demonstration of a photonic {{Aharonov}}-{{Bohm}} effect
  at radio frequencies,'' \emph{Physical Review B}, vol.~87, no.~6, p. 060301,
  Feb. 2013.

\bibitem{tzuang_non-reciprocal_2014}
\BIBentryALTinterwordspacing
L.~D. Tzuang, K.~Fang, P.~Nussenzveig, S.~Fan, and M.~Lipson,
  ``\BIBforeignlanguage{en}{Non-reciprocal phase shift induced by an effective
  magnetic flux for light},'' \emph{\BIBforeignlanguage{en}{Nature Photonics}},
  vol.~8, no.~9, pp. 701--705, Sep. 2014. [Online]. Available:
  \url{http://www.nature.com/nphoton/journal/v8/n9/full/nphoton.2014.177.html#ref8}
\BIBentrySTDinterwordspacing

\bibitem{peano_topological_2015}
\BIBentryALTinterwordspacing
V.~Peano, C.~Brendel, M.~Schmidt, and F.~Marquardt, ``Topological {Phases} of
  {Sound} and {Light},'' \emph{Physical Review X}, vol.~5, no.~3, p. 031011,
  Jul. 2015. [Online]. Available:
  \url{http://link.aps.org/doi/10.1103/PhysRevX.5.031011}
\BIBentrySTDinterwordspacing

\bibitem{andrews_quantumenabled_2015}
R.~W. Andrews, A.~P. Reed, K.~Cicak, J.~D. Teufel, and K.~W. Lehnert,
  ``\BIBforeignlanguage{en}{Quantum-enabled temporal and spectral mode
  conversion of microwave signals},'' \emph{\BIBforeignlanguage{en}{Nature
  Communications}}, vol.~6, p. 10021, Nov. 2015.

\end{thebibliography}
\end{document}